\documentstyle[aps,preprint,epsf]{revtex} \bibstyle{unsrt}

\begin{document}
\draft 


\title{Coherent Control of Atomic Beam Diffraction by Standing Light Waves} 

\author{Bijoy K. Dey} \address{Department of Chemistry, Princeton University, Princeton, New Jersey 08544}

\date{\today} \maketitle

\begin{abstract}
Quantum interference is shown to deliver a means of regulating the diffraction pattern of a thermal atomic beam interacting with two standing wave electric fields. Parameters have been identified to enhance the diffraction probability of one momentum component over the others, with specific application to Rb atoms.
\end{abstract}

\vspace*{5.0cm}
\noindent
\pacs{PACS number(s): 32.80 Qk, 34.50 Dy, 39.20 +q, 39.25 +k, 03.75 Dg}

The realization of the effect of quantum interference in manipulating and controlling atomic and molecular processes has opened up a new area of active investigation in quantum dynamics, called coherent control (CC) method \cite{coherent1}. The approach relies on the creation of a non-stationary superposition state comprising of two or more eigenstates of the atomic/molecular system. This is achieved by suitably designing an ultrashort laser pulse so that only a selective number of eigenstates are involved in the interaction between the atom and the laser pulse. Dynamics of this superposition state is entirely different from that of the eigenstate in the sense that all the superposed eigenstates contribute coherently towards the final dynamical outcome. The coherence among the eigenstates towards a particular dynamical observable can be regulated by altering the population of the eigenstates as well as their relative phases. Although this has been demonstrated lately, in the control of molecular photo-dissociation \cite{photodis}, optical racemization \cite{asymmetric}, scattering \cite{scat1}, molecular beam deposition \cite{dey1} etc., there are far fewer applications of this new innovation. In this letter, we report on the coherent manipulation of the diffraction of a thermal atomic beam by two standing light waves thus, providing a useful means towards building mirrors or beam splitters for atomic interferometers. Several works works have already been devoted to the diffraction process, \cite{diff1,control1,KD1,BRAG1} e.g., the Kapitza-Dirac diffraction of atoms \cite{KD1}, Bragg diffraction of atoms \cite{BRAG1}, etc. Classical coherence (due to the presence of two or more optical fields) and the effect of the electric field configuration in modifying the diffraction pattern has also been studied by some authors \cite{control1}. It is thus timely to introduce a different method based on the amalgamation of the classical and quantum coherence where the classical coherence (CC) is created from the phase relation between two electric field and the quantum coherence (QC) is created out of a pre-preparation of the atomic beam such that every atom in the beam lies in a superposition state of its eigenstates. The relative contribution of the CC and QC play a crucial role in regulating the atomic momenta transfer and hence, in the atomic diffraction pattern. Diffraction of an atomic beam from a light grating involves the transfer of momentum from the light field to the atoms and this depends on the excitation paths. Thus, altering excitation paths, much the way it is done in the CC of molecular photo-dissociation \cite{photodis}, would lead to the control in the atomic diffraction. Below we describe the theory and provide the computational results for Rb atom.

A thermal beam of Rb atoms is pre-prepared in a way that every atom in the beam exists in a superposition state comprising of two of its coherently related eigenstates (atomic coherence), i.e.,
\begin{eqnarray}
\Psi (t)=c_1(t)e^{-iE_1t/\hbar}|\phi _1>+c_2(t)e^{-iE_2t/\hbar}|\phi _2>
\end{eqnarray}
where $|\phi _i>$ are the eigenstates of the atomic Hamiltonian, of energy $E_i$. This preparation is achieved by passing the thermal beam through a ultra-short laser pulse decribed elsewhere \cite{dey2}. The thermal nature of the beam is characterized by its x-directional Gaussian distribution function $g(x)=(\pi \sigma _x ^2)^{-1/4}e^{-\frac{x^2}{2\sigma _x ^2}}$, where the full width at half maximum (FWHM), $2\sigma _x\sqrt{-ln(1/2)}$ of the Gaussian function directly relates to the temperature of the beam. This is followed by the passage of the beam through two coherently related standing light waves (optical coherence) whose amplitudes, frequencies and phases can be varied. That is the composit standing electric field {\bf E}(x,t) interacting with the superposed atomic state is given by
\begin{eqnarray}
{\bf E}(x,t)=\sum_jE_j(\omega _j,\theta_j,x,t)f_j(z)\hat{{\bf k}}
\end{eqnarray}
where $E_j(\omega _j,\theta _j,x,t)=E_j^{(0)}f_j(x)e^{i\omega _jt}+c.c.$ with c.c. representing the complex conjugate of the terms preceeding it, $\theta _F=|\theta _1-\theta_2|$ is the relative phase of the two standing waves (SW), $E_j^{(0)}$, $\omega _j$ and $k_j$ are the peak amplitude, frequency and wave vector of the j-th SW of wave length $\lambda _j$. This field is considered polarized along the z-direction (see Fig.1 for the general configuration of the proposed scenario). The functions $f_j(x)$ and $f_j(z)$ are the light beam profiles of the j-th electric field along the x and z directions respectively. We assume the atomic momentum along z is sufficiently larger than that along x so that light forces along z can be neglected and the longitudinal velocity $v_z$ remains constant. This effectively means that we can replace z in $f_j(z)$ by $z = z_{in}+v_zt$, where $z_{in}$ (set as zero) is the nozzle position where the beam ejects from. Taking $f_1(z)=f_2(z)=exp(-(z-z_{0})^2/\sigma _z^2)$, we replace it by $f(t)=exp(-(t-t_0)^2/\tau ^2)$, where $\tau = \sigma _z/v_z$ and $t_0=z_{0}/v_z$ represent the interaction region and the time when the SWs have the maximum intensity. Thus $f_j(z)$ represents the adiabatic entry and exit for the atomic beam. The x-dependence $f_j(x)$ is taken as $f_j(x)=2cos(k_jx+\theta _j)$ which means that j-th electric field consists of two components which counterpropagate each other.

The excitation now takes place from the superposition state to the higher excited states of the atom (see Fig.1 for the atomic transitions driven by the SW fields). We assume that the interaction time is sufficiently small so that the transverse kinetic energy absorbed by the atoms during the interaction can be neglected (Raman-Nath approximation) and also that the freqencies of the light fields are well detuned from the atomic resonances so that spontaneous emission can be neglected. Thus we have an effective three-level system interacting with the field given by Eq.(2). Within rotating wave approximation (RWA) the Schr\"{o}dinger equation for this system is

\begin{eqnarray}
\left(\begin{array}{c}\dot{a_1} \\ \dot{a_2} \\ \dot{a_3}
\end{array}\right)=\left(\begin{array}{ccc}0 & if(t)\sum_{j=1}^2\Omega
^{(1,2)}_jf_je^{-i\Delta ^{(1,2)}_jt} & if(t)\sum _{j=1}^2\Omega
^{(1,3)}_jf_je^{-i\Delta ^{(1,3)}_jt}\\ if(t)\sum _{j=1}^2\Omega
^{(1,2)}_jf_je^{i\Delta ^{(1,2)}_jt} & 0 &
if(t)\sum_{j=1}^2\Omega^{(2,3)}_jf_je^{-i\Delta ^{(2,3)}_jt} \\ if(t)\sum
_{j=1}^2\Omega ^{(1,3)}_jf_je^{i\Delta ^{(1,3)}_jt} & if(t)\sum_{j=1}^2\Omega
^{(2,3)}_jf_je^{i\Delta ^{(2,3)}_jt} & 0
\end{array}\right)\left(\begin{array}{c}a_1 \\ a_2 \\ a_3
\end{array}\right)
\end{eqnarray}
where $\Omega _j^{m,n}=<\phi _m|\stackrel{\rightarrow}{\mu}|\phi _n>E_j^{(0)}/\hbar$ is the Rabi frequency, $\Delta _j^{m,n}=1/\hbar(E_n-E_m)-\omega _j$ is the detuning of the j-th SW field and $\stackrel{\rightarrow}{\mu}$ the atomic dipole vector. Integration of the above equation followed by the Fourier transformation (FT) delivers $\tilde {a}_1(k)$, $\tilde{a}_2(k)$ and $\tilde{a}_3(k)$, where tilda refers to the FT of the respective time function. The functions $\tilde {a}_1(k)$, $\tilde{a}_2(k)$ and $\tilde{a}_3(k)$ jointly describe the diffraction pattern of the atomic beam. Within second order perturbative theory we obtain $\tilde {a}_1(k)$ and $\tilde {a}_2(k)$ as
\begin{eqnarray}
\tilde{a}_1(k)&=&\tilde{g}(k)c_1+\frac{ic_2}{2}\sum_{j,s}\Omega ^{(1,2)}_jI_1^{(j)}e^{si\theta _j}\tilde{g}(k-sk_j)-\frac{1}{4}\sum _{j,j^\prime,s}[c_1\Omega _j^{(1,2)}\Omega _{j^{\prime}}^{(1,2)}I_2^{(j,j^{\prime})}\\\nonumber
&&+c_1\Omega _j^{(2,3)}\Omega _{j^{\prime}}^{(1,3)}I_3^{(j,j^{\prime})}+c_2\Omega _j^{(1,3)}\Omega _{j^{\prime}}^{(2,3)}I_4^{(j,j^{\prime})}][e^{si(\theta _j+\theta _{j^{\prime}})}\tilde{g}(k-s(k_j+k_{j^{\prime}}))\\\nonumber
&&+e^{si(\theta _j-\theta _{j^{\prime}})}\tilde{g}(k-s(k_j-k_{j^{\prime}}))]
\end{eqnarray}

\begin{eqnarray}
\tilde{a}_2(k)&=&\tilde{g}(k)c_2+\frac{ic_1}{2}\sum_{j,s}\Omega ^{(1,2)}_jI_1^{(j)^{*}}e^{si\theta _j}\tilde{g}(k-sk_j)-\frac{1}{4}\sum _{j,j^\prime,s}[c_2\Omega _j^{(1,2)}\Omega _{j^{\prime}}^{(1,2)}I_2^{(j,j^{\prime})^{*}}\\\nonumber
&&+c_2\Omega _j^{(2,3)}\Omega _{j^{\prime}}^{(2,3)}I_5^{(j,j^{\prime})}+c_1\Omega _j^{(2,3)}\Omega _{j^{\prime}}^{(1,3)}I_4^{(j,j^{\prime})^{*}}][e^{si(\theta _j+\theta _{j^{\prime}})}\tilde{g}(k-s(k_j+k_{j^{\prime}}))\\\nonumber
&&+e^{si(\theta _j-\theta _{j^{\prime}})}\tilde{g}(k-s(k_j-k_{j^{\prime}}))]
\end{eqnarray}
where the initial conditions ($\tau=0$) are $\tilde {a}_1(k)=c_1\tilde{g}(k)$, $\tilde {a}_2(k)=c_2\tilde{g}(k)$ and $\tilde{a}_3(k)=0$. If the intensity of the SW fields are relatively low and/or the detunings from the resonances are relatively large, the excitation from  the initial pre-prepared state given by Eq.(1) would be constrained significantly and the contribution of $\tilde{a}_3(k)$ would remain negligible. This is often referred to as adiabatic evolution of the initial state \cite{ADAMS}. In such case, the light-atom interaction is equivalent to the propagation of a scalar atomic wave through an optical potential and the deflection of the atomic de Broglie waves by light fields is exactly analogous to that of the light by a dielectric medium \cite{ADAMS}. The summation index s has two values viz., +ve and -ve, whereas $I_1^{(j)}$, $I_2^{(j,j^{\prime})}$, $I_3^{(j,j^{\prime})}$, $I_4^{(j,j^{\prime})}$ and $I_5^{(j,j^{\prime})}$ are given by
\begin{eqnarray*}
I_1^{(j)}=\int_{-\infty}^{\infty}f(t)e^{-it\Delta_j^{(1,2)}}dt
\end{eqnarray*}
\begin{eqnarray*}
I_2^{(j,j^{\prime})}=\int_{-\infty}^{\infty}dt^\prime f(t^\prime)e^{-it^{\prime}\Delta_j^{(1,2)}}\int_{-\infty}^{t^{\prime}}f(t^{\prime \prime})e^{it^{\prime \prime}\Delta_{j^{\prime}}^{(1,2)}}dt^{\prime \prime}
\end{eqnarray*}
\begin{eqnarray*}
I_3^{(j,j^{\prime})}=\int_{-\infty}^{\infty}dt^\prime f(t^\prime)e^{-it^{\prime}\Delta_j^{(1,3)}}\int_{-\infty}^{t^{\prime}}f(t^{\prime \prime})e^{it^{\prime \prime}\Delta_{j^{\prime}}^{(1,3)}}dt^{\prime \prime}
\end{eqnarray*}
\begin{eqnarray*}
I_4^{(j,j^{\prime})}=\int_{-\infty}^{\infty}dt^\prime f(t^\prime)e^{-it^{\prime}\Delta_j^{(1,3)}}\int_{-\infty}^{t^{\prime}}f(t^{\prime \prime})e^{it^{\prime \prime}\Delta_{j^{\prime}}^{(2,3)}}dt^{\prime \prime}
\end{eqnarray*}
and
\begin{eqnarray*}
I_5^{(j,j^{\prime})}=\int_{-\infty}^{\infty}dt^\prime f(t^\prime)e^{-it^{\prime}\Delta_j^{(2,3)}}\int_{-\infty}^{t^{\prime}}f(t^{\prime \prime})e^{it^{\prime \prime}\Delta_{j^{\prime}}^{(2,3)}}dt^{\prime \prime}
\end{eqnarray*}
and * represents complex conjugate. The function $\tilde{g}(k)$ is the Fourier transform of g(x). The scattered wave function of the atom is a superposition of the Gaussian modulated plane waves with momenta $\hbar k$=0, $\pm \hbar k_1$, $\pm \hbar k_2$, $\pm \hbar (k_1-k_2)$, $\pm \hbar (k_1+k_2)$, $\pm 2\hbar k_1$ and $\pm 2\hbar k_2$. Thus, in the coherent control method the momentum transfered from the field to the atom has components different than when there was no coherence. The momentum $\hbar (k_1+k_2)$ $(\hbar (k_1-k_2))$ corresponds to the absorption of a photon from the $+k_1$ component of the SW followed by induced emission in the $-k_2$ $(k_2)$ component of the SW and is the result of the optical coherence. 

The output diffraction probability which can be measured experimentally, given by $|\tilde{a}_1(k)+\tilde{a}_2(k)|^2$ represents a comb of images of the split atomic velocities $v=\hbar k/m$, where m is the mass of the atom. This output depends on the controlled parameters $c_1$, $c_2$, $\theta _M$, $\theta _F$, $\tau$, $\sigma _x$, $E_1^{(0)}$, $E_2^{(0)}$ and the superposed states, $|\phi _1>$ and $|\phi _2>$ with $\theta_M$ being the atomic phase which is the relative phase between $c_1$ and $c_2$. The superposition is created between $|n,l_1,m_1>$ and $|n,l_2,m_2>$ Rydberg states employing a two photon preparatory stage (see ref.\cite{dey2}), where n refers to the principal quantum number of the ground state of Rb, $l_i$ and $m_i$ are the angular and azimuthal quantum numbers respectively. The Rydberg states were evaluated following the quantum defect theory. Thus, the process of excitation to the state $|\phi _3>$ occurs through three different paths viz., (1) the path that ends at $|\phi _3>$ from $|n,l_1,m_1>$, the probability of which is proportional to $|c_1|^2$, (2) the path that ends at $|\phi _3>$ from $|n,l_2,m_2>$, the probability of which is proportional to $|c_2|^2$ and (3) the path that ends at $|\phi _3>$ through the interference between the paths (1) and (2), the probability of which is proportional to $|c_1c_2|cos \theta _M$, where $\theta_M$ can be controlled during the preparation of the superposition state. We chose the frequency of the SW fields as $\omega _j=Min\{|E_2-E_1|/\hbar,|E_3-E_1|/\hbar,|E_3-E_2|/\hbar\}+\Delta _j$ which gives the detunings defined in Eq.(3) as $\Delta_j^{m,n}=1/\hbar(E_n-E_m)-Min\{|E_2-E_1|/\hbar,|E_3-E_1|/\hbar,|E_3-E_2|/\hbar\}-\Delta _j$ where m and n run from 1 to 3. In the calculation below for Rb atomic beam we chose $\Delta _1=53051.6$ $cm^{-1}$ and $\Delta _2=477464.8$ $cm^{-1}$ which ensure adiabatic evolution of the initial state. This means that the first and the second SW fields are  detuned by $\Delta _1$ and $\Delta _2$ respectively from the minimum of the atomic resonances.

As an example, we examine the diffraction where the beam is peaked about zero along the transverse direction with a FWHM=5.5 $\mu m$ which means the initial transverse spread (FWHM) of the atomic wave vector FWHM=0.504 $\mu m^{-1}$. This gives high resolution in the diffraction pattern of Rb beam as shown in Fig.2 for $|\tilde{a}_1(k)|^2$ (label a), $|\tilde{a}_2(k)|^2$ (label b), $|\tilde{a}_1(k)|^2+|\tilde{a}_2(k)|^2$ (label c) and $|\tilde{a}_1(k)+\tilde{a}_2(k)|^2$ (label d). Time integration is performed over a time period  $t=-5\times FWHM_t$ to $t=5\times FWHM_t$, where $FWHM_t=2\tau \sqrt{-ln(1/2)}$ is the full width at half maximum of the gaussian time profile centered at $t_0=0$. In the computation we have taken $FWHM_t=5 \mu s$ which means that the superposed atomic beam interacts with the standing wave fields for a total duration of 50 $\mu s$ \cite{footnote1}, during which the fields become maximum at t=25 $\mu s$. In other words, for a fixed longitudinal velocity \cite{footnote2} $v_z=500 m/s$ the field is maximum at distant of 12.5 mm from the nozzle. In Fig.2 we observed the splitting of the beam into a total of 9 velocity components, of which the k=0 is the most peaked. For the standing wave fields with $\lambda _1=0.897 \mu m$ ($k_1=7.004 \mu m^{-1}$) and $\lambda _2=1.45 \mu m$ ($k_2=4.338 \mu m^{-1}$) the beam splits at k=0, 11.332, -11.332, 2.666, -2.666, 14.008, -14.008, 8.676 and -8.676 $\mu m^{-1}$. We observed that the incorporation of more coherence (optical) through the introduction of more SW fields will give rise to the splitting of the atomic beam into a wide range of velocity components. Qualitative consideration of the pattern reveals that they are in general agreement with the theory \cite{theory}. That is, the atom absorbs photon from the $+ k_j$ components followed by induced emission in the $-k_{j^{\prime}}$ components of the standing wave where j and $j^{\prime}$ can be 1 and/or 2. 

We compute the relative diffraction probabilities of any two different momenta components $P_{k^{\prime}k^{\prime \prime}}=P(k^{\prime})/P(k^{\prime \prime})$ where $P(k^{\prime})$ and $P(k^{\prime \prime})$ are the total diffraction probabilities corresponding to the atomic momenta $\hbar k^{\prime}$ and $\hbar k^{\prime \prime}$ respectively. This $P(k^{\prime})$ is computed by filtering out all the momenta components in the diffraction profile excepting the one centered at $k^{\prime}$ and then integrating over the entire momentum coordinate, i.e., $P(k^{\prime})=\int |\tilde{a}_1(k)+\tilde{a}_2(k)|^2 F(k,k^{\prime})dk$, where $F(k,k^{\prime})$ is the filter chosen properly. Consideration of the equations for $\tilde{a}_1(k)$ and $\tilde{a}_2(k)$ (Eqs. (4) and (5)) show that the changes in the control parameters can strongly affect $P_{k^{\prime}k^{\prime \prime}}$, in other words, we can control the degree of splitting of one of the atomic velocity component over the others. The following figures (Fig.3-5) show the dependence of $P_{k^{\prime}k^{\prime \prime}}$ on atomic phase $\theta_M$ (Fig.3), $|c_1|^2/|c_2|^2$ (Fig.4) and the optical phase $\theta _F$ (Fig.5). All these figures show significant control of the diffraction probability of a given momentum component over the others by varying the controlled parameters. For example, Fig.3 (label c) show that the diffraction probability of k=0 component of the atomic wave vector is nearly 58 times that of the $k=2k_2$ component at $\theta _M=0.57$ radian, i.e., $P(0)\approx 58 P(2k_2)$. This gets altered at $\theta _M=3.74$ radian when the diffraction probability of the $k=2k_2$ component become nearly 11 times that of the k=0 component. Take another example, Fig.4 (label j), where we see that the diffraction probability of $k=2k_2$ is roughly 12 times that of $k=2k_1$ for $\theta _M=-3.2$ radian which totally reverses in favor of the $k=2k_1$ component at $\theta _M=-0.45$ radian when $P(2k_1)\approx 11 P(2k_2)$. Thus the atomic phase causes a significant control on the diffraction probabilities of the atomic beam for fixed $|c_1|^2$ and $|c_2|^2$ (Fig.3).

Consider now the control over the Rb beam where initially $|c_1|^2 \times 100$ precentage of the atoms lie in the state $|5,0,0>$ and the rest in the state $|5,2,0>$. Results for such case are shown in Figs.4 where we plot the logarithms of the relative diffraction probabilities of two different atomic wave vector components for varying $|c_1|^2/|c_2|^2$. Results show near monotonic decrease of $P(0)/P(k_1-k_2)$, $P(0)/P(k_1+k_2)$, $P(0)/P(2k_2)$ and $P(0)/P(2k_1)$ with the increase in $|c_1|^2/|c_2|^2$ for up to $|c_1|^2/|c_2|^2\approx 3.5$, after which they become nearly constant (Fig.4a). Whereas $P(k_1-k_2)/P(k_1+k_2)$, $P(k_1-k_2)/P(2k_2)$, $P(k_1-k_2)/P(2k_1)$, $P(k_1+k_2)/P(2k_2)$, $P(k_1+k_2)/P(2k_1)$, $P(2k_2)/P(2k_1)$ increase monotonically with the increase in $|c_1|^2/|c_2|^2$ for up to $|c_1|^2/|c_2|^2\approx 3.5$ and then become constant (Fig.4 b, c, d). Finally, Fig.5 depicts the control of the relative diffraction probabilities of two different atomic wave vectors for varying optical phase $\theta _F$, where we again observe several-fold increase (or decrease) of diffraction probabilities of a particular atomic wave vector over the other.

The present work takes into account the experimental issues, e.g., (1) the experimental noise is taken care of by considering a gaussian thermal beam and (2) the entry and the exit of the atomic beam in its interaction with the SW fields is described by a gaussian envelope in the calculation. The parameters in the present calculation have been chosen carefully so as to commensurate with the Rb atom and the assumptions that the spontaneous emission is negligible and Raman-Nath regime is valid. These assumptions do not pose any obstacle in the real experimental scenario, they mere confine the intensity and the frequency of the SW fields within certain values.

In conclusion, we have shown that one can achieve significant control on the diffraction probability of a thermal atomic beam through the introduction of the atomic and optical coherences. Although we have introduced atomic coherence by two-state superposition scenario, a more general approach should be the preparation of the superposition state by an arbitrary numbers of atomic eigenstates accessible to the preparatory electric field. This would immediately introduce more parameters and hence more control over the diffraction pattern.
 
\vspace{0.2in}
{\bf Acknowledgement}
The author wishes to thank Prof. Paul Brumer for helpful discussions. This research was performed in part using resources at the Chemistry Department of the University of Toronto.

\pagebreak

{\bf Figure Captions}

Figure 1: Schematic of proposed control scenario.

\vspace{0.2in}
Figure 2: Atomic diffraction pattern associated with the initial superposition states $|\phi _1>=|5,0,0>$ and $|\phi _2>=|5,2,0>$. Here $|c_1|^2=0.8$, $|c_2|^2=0.2$, $E_1^{(0)}=1\times 10^4$ V/m, $E_2^{(0)}=1\times 10^4$ V/m, $FWHM_x=5.5 \mu m$, $FWHM_t=5.0 \mu s$, $\theta_F=4.0$ radian, $\theta_M$=4.6 radian, $\lambda_1=0.897 \mu m$, $\lambda_2=1.45 \mu m$ and $t_0=0$. The lebels 1, 2, 3, 4, 5, 6, 7, 8 and 9 correspond to the atomic momenta 0, $\hbar (k_1-k_2)$, $-\hbar (k_1-k_2)$, $-2 \hbar k_2$, $-\hbar (k_1+k_2)$, $-2 \hbar k_1$, $2\hbar k_2$, $\hbar (k_1+k_2)$ and $2 \hbar k_1$ respectively. The curves are normalized.

\vspace{0.2in}
Figure 3: Relative integrated diffraction probabilities $P_{k^{\prime}k^{\prime \prime}}$corresponding to the atomic momenta (a) $\hbar k^{\prime}=0$, $\hbar k^{\prime \prime}=k_1-k_2$; (b) $\hbar k^{\prime}=0$, $\hbar k^{\prime \prime}=k_1+k_2$; (c) $\hbar k^{\prime}=0$, $\hbar k^{\prime \prime}=2k_2$; (d) $\hbar k^{\prime}=0$, $\hbar k^{\prime \prime}=2k_1$; (e) $\hbar k^{\prime}=k_1-k_2$, $\hbar k^{\prime \prime}=k_1+k_2$; (f) $\hbar k^{\prime}=k_1-k_2$, $\hbar k^{\prime \prime}=2k_2$; (g) $\hbar k^{\prime}=k_1-k_2$, $\hbar k^{\prime \prime}=2k_1$; (h) $\hbar k^{\prime}=k_1+k_2$, $\hbar k^{\prime \prime}=2k_2$; (i) $\hbar k^{\prime}=k_1+k_2$, $\hbar k^{\prime \prime}=2k_1$; (j) $\hbar k^{\prime}=2k_2$, $\hbar k^{\prime \prime}=2k_1$; for the initial superposition states $|\phi _1>=|5,0,0>$ and $|\phi _2>=|5,2,0>$ plotted against $\theta _M$. Other parameters are as in Fig.2.

\vspace{0.2in}
Figure 4: Logarithm of the relative integrated diffraction probabilities $ln(P_{k^{\prime}k^{\prime \prime}})$ plotted against $|c_1|^2/|c_2|^2$ for the initial superposition states $|\phi _1>=|5,0,0>$ and $|\phi _2>=|5,2,0>$. Here the lebels a, b, c, d, e, f, g, h, i, j signify the same as in Fig.2. Other parameters are as in Fig.2.

\vspace{0.2in}
Figure 5: Relative integrated diffraction probabilities $P_{k^{\prime}k^{\prime \prime}}$ plotted against $\theta _F$ for the initial superposition states $|\phi _1>=|5,0,0>$ and $|\phi _2>=|5,2,0>$. Here the lebels a, b, c, d, e, f, g, h, i, j signify the same as in Fig.2. Other parameters are as in Fig.2.

\begin{figure}[!h]
\epsfxsize=6.0in
\hspace*{0.5cm}\epsffile{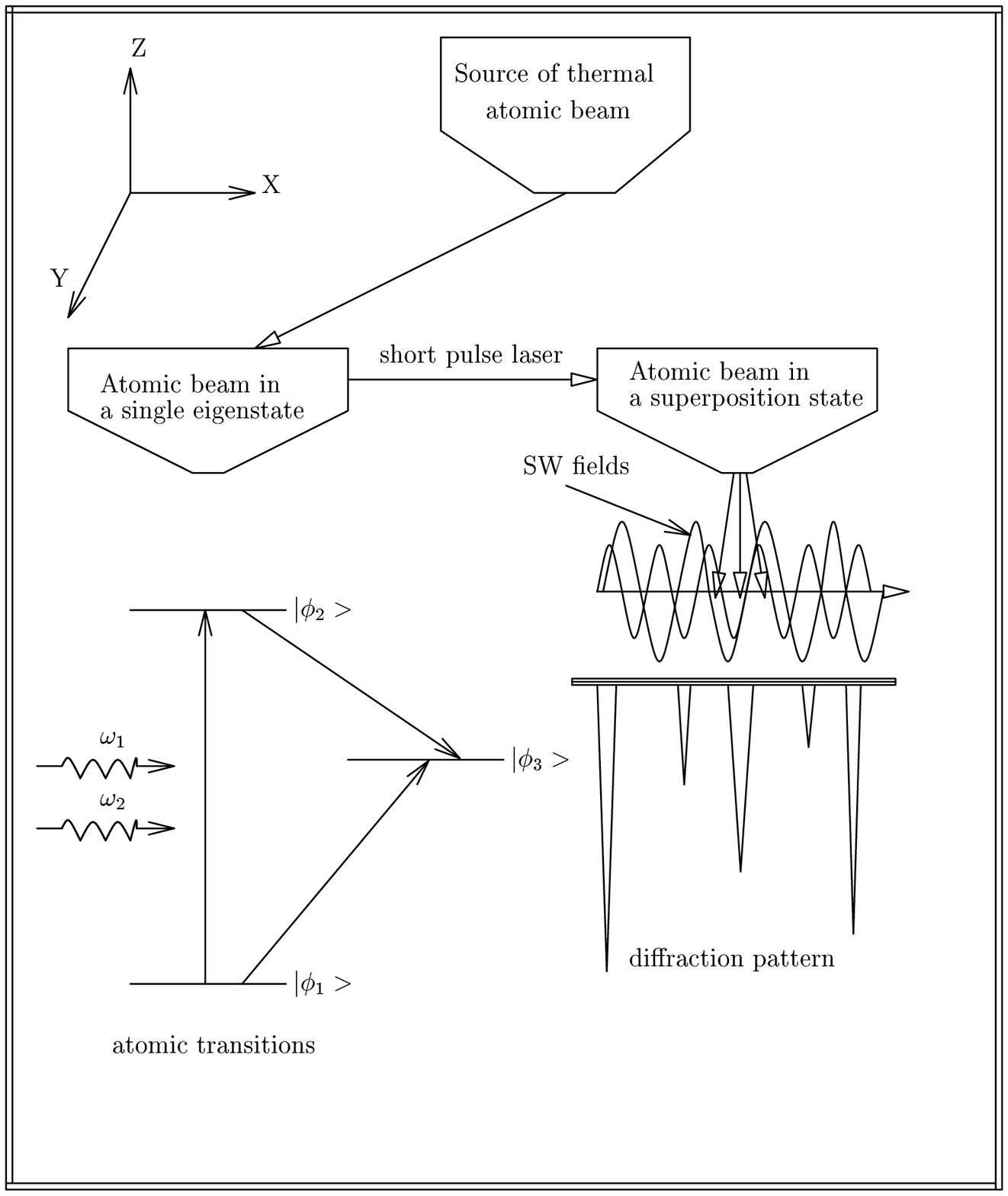}
\vspace*{-0.5cm}\caption{} 
\label{diff_z}
\end{figure}

\begin{figure}[!h]
\epsfxsize=6.0in
\hspace*{0.5cm}\epsffile{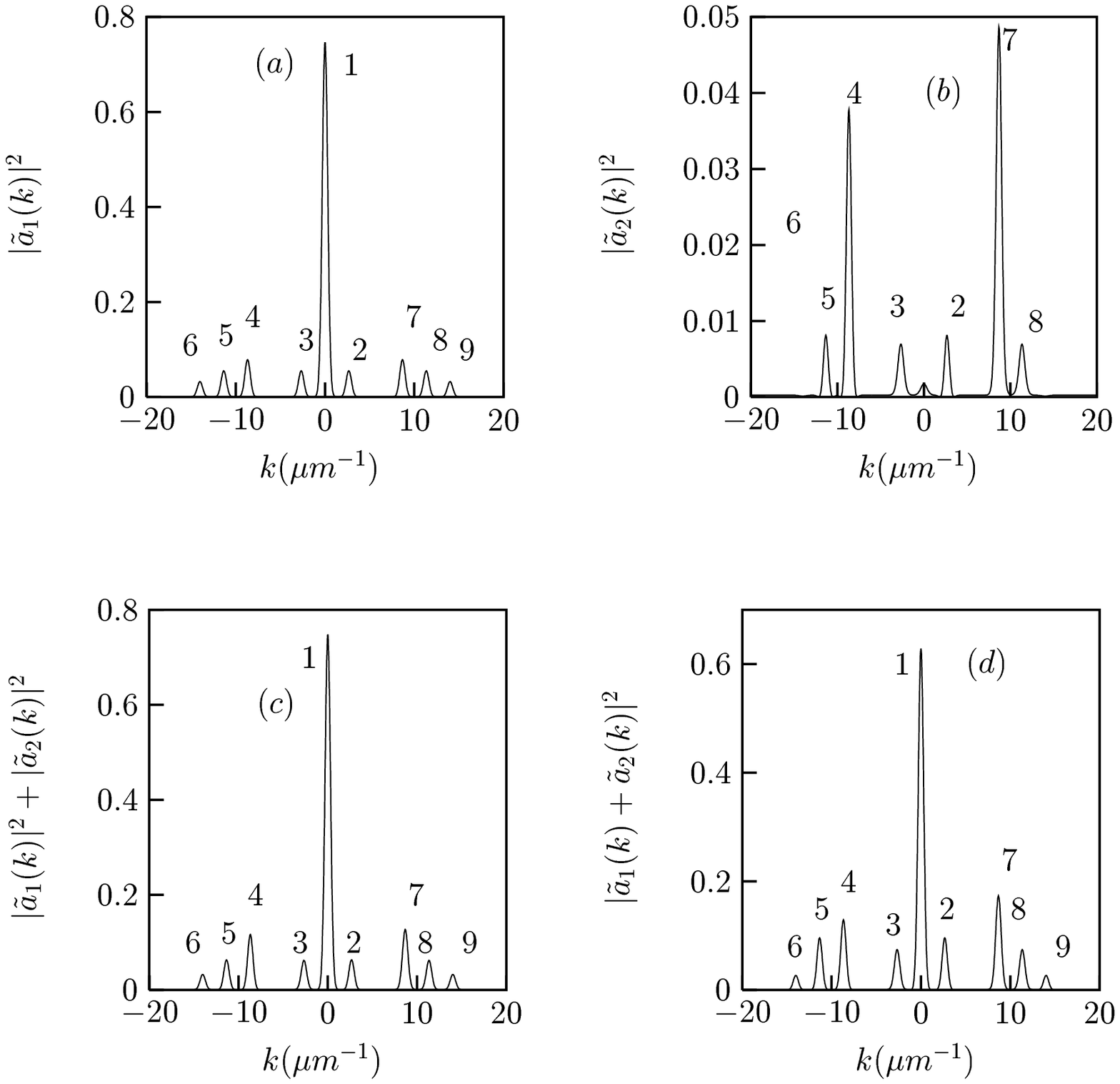}
\vspace*{-0.5cm}\caption{} 
\label{diff_z}
\end{figure} 

\begin{figure}[!h]
\epsfxsize=6.0in
\hspace*{0.5cm}\epsffile{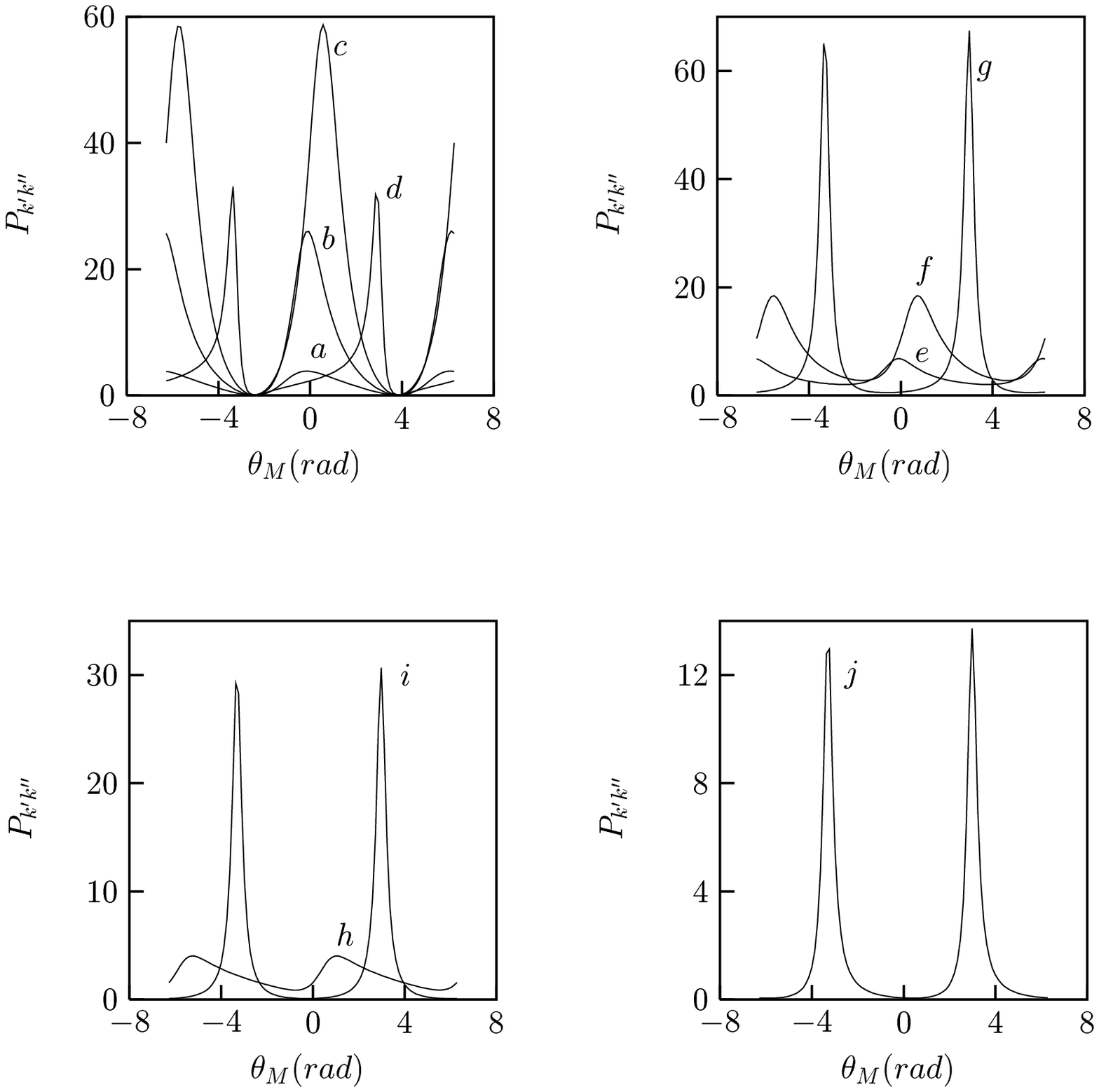}
\vspace*{-0.5cm}\caption{}
\label{diff_thetaM}
\end{figure}

\begin{figure}[!h]
\epsfxsize=6.0in
\hspace*{0.5cm}\epsffile{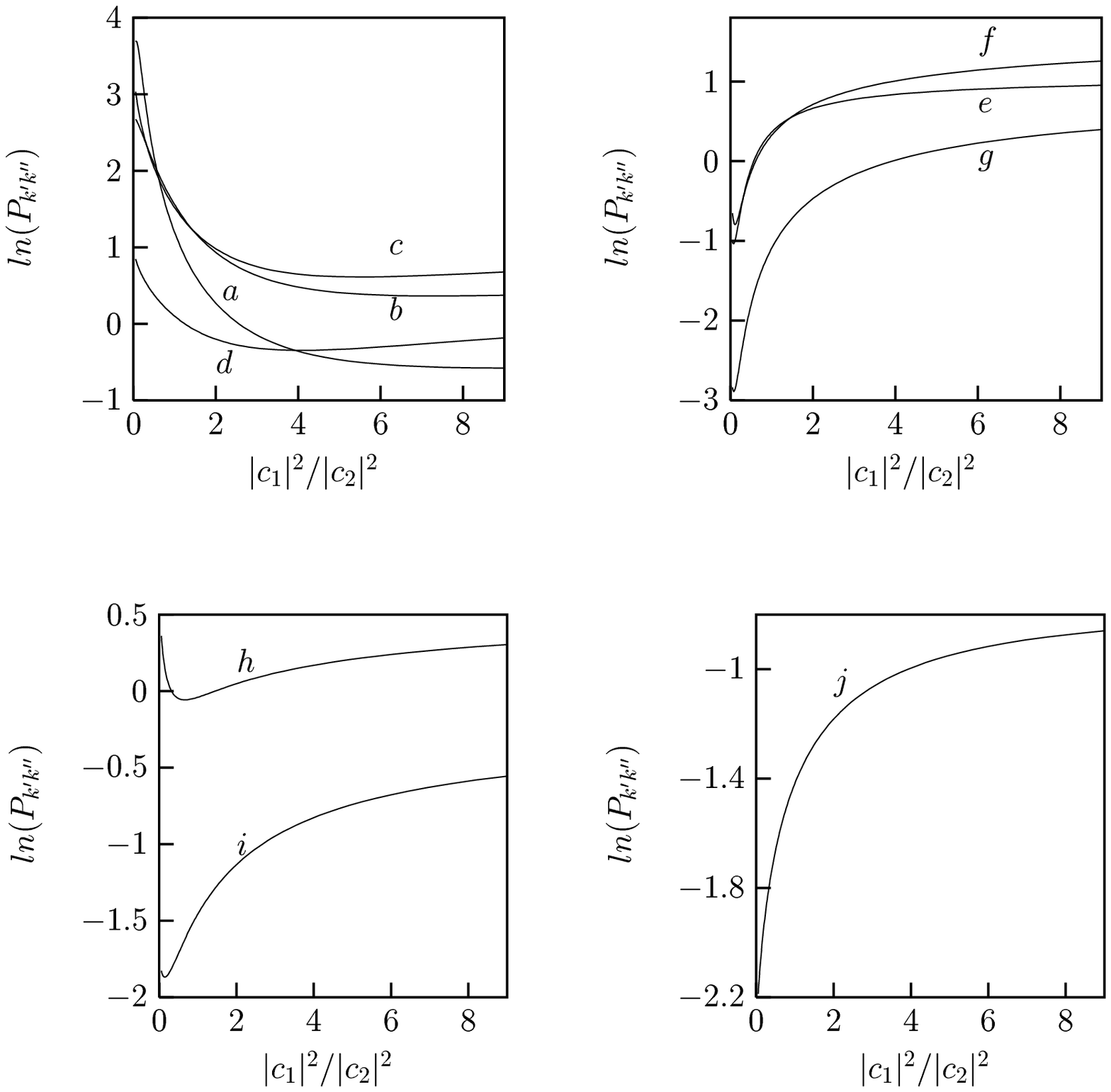}
\vspace*{-0.5cm}\caption{}
\label{diff_c12}
\end{figure}

\begin{figure}[!h]
\epsfxsize=6.0in
\hspace*{0.5cm}\epsffile{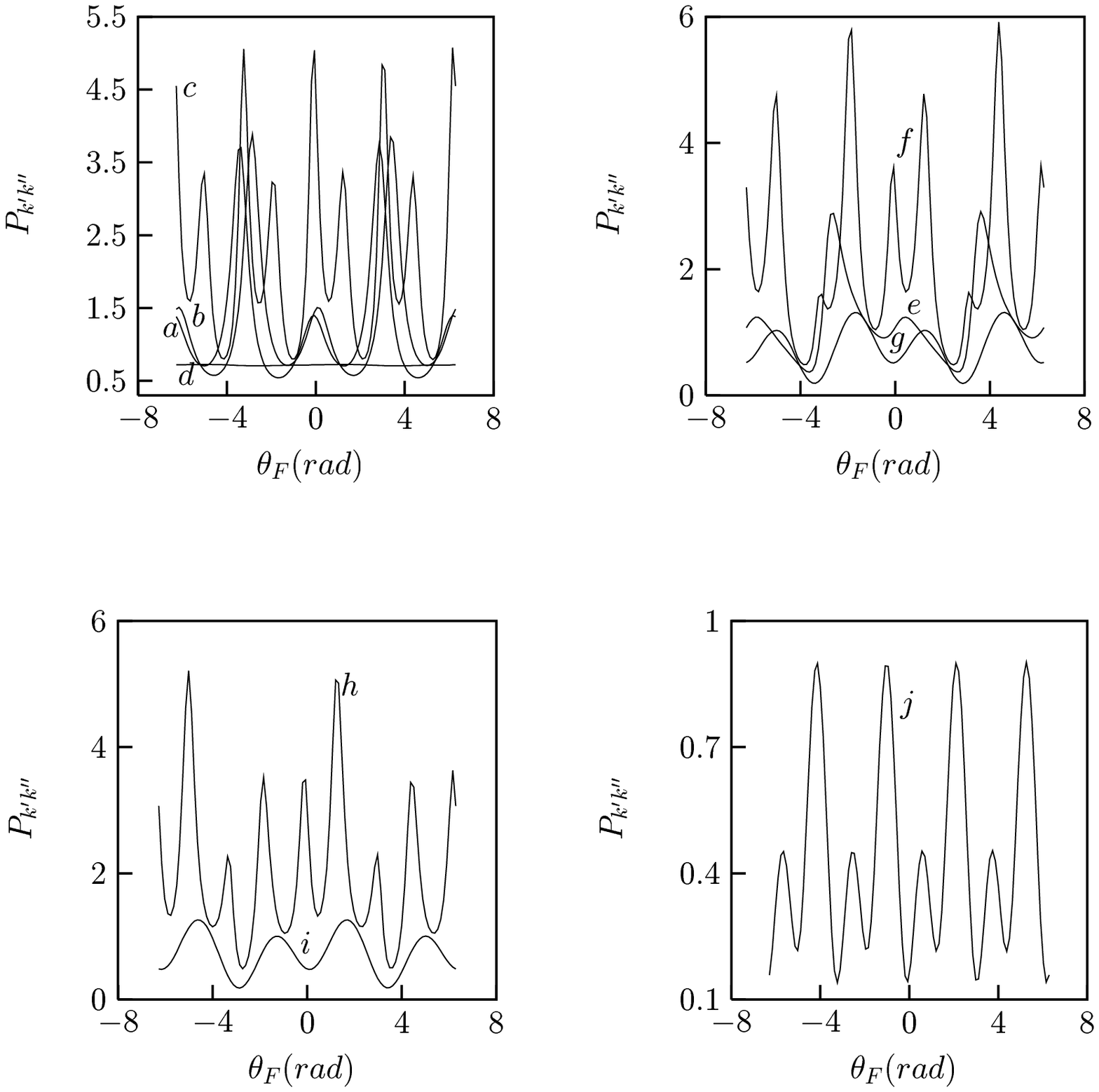}
\vspace*{-0.5cm}\caption{}
\label{diff_thetaF}
\end{figure}


\begin{references}
\bibitem{footnote1}Time of interaction plays a crucial role in the sense that $\tau $must be less that T/4, where T is the optical period of the atom in the potential well $V(x)=-1/2\alpha E^2(x)$. Here $\alpha $ is the polarizability of the Rb atom and E(x) is the electric field with the time variable integrated out (see ref.\cite{dey1}).
\bibitem{footnote2}More detail study would require assigning the beam a proper transverse velocity profile corresponding to its transverse temperature. However, this does not alter the results in the present computation because of the fact that the atoms do not experience any force along the transverse direction.
\end{references}
\end{document}